\documentstyle[11pt,epsf]{article}
\newcommand{\al}{\alpha}

\newcommand{\bt}{\beta}
\newcommand{\bv}{\begin{verbatim}}
\newcommand{\de}{\delta}

\newcommand{\De}{\Delta}

\newcommand{\ep}{\epsilon}

\newcommand{\ev}{\end{verbatim}}
\newcommand{\fr}{\frac}

\newcommand{\ga}{\gamma}

\newcommand{\hb}{\hbar}

\newcommand{\la}{\lambda}

\newcommand{\na}{\nabla}

\newcommand{\ph}{\phi}
\newcommand{\ps}{\psi}

\newcommand{\si}{\sigma}
\newcommand{\ta}{\tau}

\newcommand{\vb}{\verb}

\newcommand{\be}{\begin{equation}}
\newcommand{\ee}{\end{equation}} 
\newcommand{\eei}{\end{equation}\indent\indent}
\newcommand{\bc}{\begin{center}}
\newcommand{\ec}{\end{center}}
\newcommand{\ber}{\begin{eqnarray}}
\newcommand{\ear}{\end{eqnarray}}
\newcommand{\ba}{\begin{array}}
\newcommand{\ea}{\end{array}}

\newcommand{\p}{\partial}

\def\case#1/#2{\textstyle\frac{#1}{#2} }
\begin{document}
\title{The Quantization of Geodesic Deviation.}
\author{Mark D. Roberts, \\\\
Department of Mathematics and Applied Mathematics, \\ 
University of Cape Town,\\
Rondbosch 7701,\\
South Africa\\\\
roberts@gmunu.mth.uct.ac.za} 
\date{\today}
\maketitle
\vspace{0.1truein}
\bc Published:  {\it Gen.Rel.Grav.} {\bf 28}(1996)1385-1392.\ec
\bc Eprint: gr-qc/9903097\ec
\bc Comments:  9 pages, no diagrams,  no tables,  Latex2e.\ec
\bc 3 KEYWORDS:\ec
\bc Geodesic Deviation:~~  
Spreading of the Wave Packet:~~  
one to many particle interpretation.\ec
\bc 1999 PACS Classification Scheme:\ec
\bc http://publish.aps.org/eprint/gateway/pacslist \ec
\bc 11.15q,  03.70+k,  04.60+n.\ec
\bc 1991 Mathematics Subject Classification:\ec
\bc http://www.ams.org/msc \ec
\bc 81T13,  83C10,  81T20,  83C45.\ec
\newpage
\begin{abstract}
There exists a two parameter action,   
the variation of which produces both the geodesic 
equation and the geodesic deviation equation.   
In this paper it is shown that this action can be 
quantized by the canonical method,  resulting in equations 
which generalize the Klein-Gordon equation.   The resulting
equations might have applications,  and also show that
entirely unexpected systems can be quantized.   The possible 
applications of quantized geodesic deviation are to:
i)the spreading wave packet in quantum theory,  ii)and also 
to the one particle to many particle problem in second 
quantized quantum field theory.
\end{abstract}
\section{Introduction.}
\label{sec:intro}
In relativity the path of a particle which only interacts with the 
gravitational field is given by a geodesic,  and this can be quantized 
to give the Klein-Gordon equation.   The paths of many particles which 
only interact with the gravitational field are described by the geodesic 
deviation equation.  Both the geodesic and geodesic deviation equations 
have been derived from simultaneous actions,  
Bazanski (1977) \cite{bi:baz77}.   How to 
quantize such systems is not immediately apparent as it is not clear which 
action to start with.   The coordinate space action which produces the general 
geodesic and geodesic deviation equations has non-covariant associated momenta 
which depend on the Christoffel symbol and hence the coordinates. Quantization 
of a non-covariant system is not acceptable because the wave function would 
depend on the coordinates used.   Instead we use a coordinate space action 
which has unit normalized momenta $p_\al p^\al_.=1$ and is covariant;  
variations of the corresponding phase space action give the full 
non-normalized geodesic and geodesic deviation equations and thus describe 
the whole system.  Quantization of the phase space action is carried out by 
the canonical method, Dirac (1963) \cite{bi:dirac}.

There are at least three motives for studying the quantization of such 
actions,  only the first is touched on here.   The first is that the actions 
are gauge invariant systems  with two gauge parameters,  and as such provide 
a testing ground for the techniques for quantizing gauge systems.

The second is that in quantum theory the wave packet spreads, whereas in 
gravitational theory matter attracts.   In Riemann geometry the Ricci identity 
relates the Riemann tensor to the commuted covariant derivatives of a vector 
field.   The covariant derivative of a vector field can be decomposed into 
acceleration , expansion,  shear and vorticity.   If this decomposition is 
inserted into the Ricci identity and then the resulting equations contracted 
and transvected one obtains Raychaudhuri's equation,  Hawking and Ellis 
(1973) \cite{bi:HE}p.84.   Furthermore if an energy condition 
(the time-like convergence condition) is assumed,  
then all the terms (except the vorticity) show that 
the rate of expansion is negative:  in other words the effect of gravitation 
is to make a fluid contract or a system of particles coalesce.  On the other
hand in non-relativistic quantum mechanics ,as described by Schr\"odinger's
equation,  see for example Schiff (1968) \cite{bi:schiff}p.64-65;  
a wave packet can be constructed which is taken to correspond to a 
free non-interacting classical particle.   As time increases the wave 
packet spreads.   Thus for a system of particles described by classical 
gravitation there is a contraction of the system as opposed to a 
quantum non-relativistic single particle system where there is expansion.   
Now a quantization of geodesic deviation describes
both many particles and is a quantum system - so would there be overall 
expansion or contraction of the wave function?   There is a possibility that 
the two effects cancel out.

The third is that the Klein-Gordon equation arises from the quantization 
of a single particle,  and in quantum field theory second quantization of this
produces a many particle theory,  Itzykson and Zuber (1985) \cite{bi:IZ}p.110;
from a physical point of view it would be more consistent to go from a many 
particle theory to a many particle theory.   When this is done the wave 
function depends upon both $x$ and $r$,  and so takes account of both the 
positions of the particles and there relative motion.   
In the approach used here assessment of the contribution of $x$ and $r$ 
to the wave function is hampered because the wave
function is not separable into $x$ and $r$ dependent parts.   There are higher 
deviation equations Bazanski (1977) \cite{bi:baz77} and their quantization 
can be done by the same methods.   All the deviation equations can be 
produced from a Taylor series expansion Bazanski (1976) \cite{bi:baz76},  
and it might be that there is a single wave 
equation which incorporates the wave equation corresponding to all the 
deviation equations.   The conventions used are:  signature $+---$,  Riemann 
tensor $-2X_{\al;[\bt\ga]}=X^\de R_{\de\al\bt\ga}$, $D$ and $\na$ 
signify covariant differentiation.
\section{The Coordinate Space Action.}
\label{sec:2}
The simultaneous dynamic coordinate space action is discussed in
Bazanski (1977) \cite{bi:baz77} eq.2.28 and is
\be
2W=\int^{\ta_1}_{\ta_0}u_\al\fr{Dr^\al}{d\ta}d\ta.
\label{eq:2.1}
\ee
The gauge invariances are
\be
\ta\rightarrow\bar{\ta}=\ta+\ep_1,~~~
r^\mu\rightarrow\bar{r}^\mu=r^\mu+\ep_2u^mu.
\label{eq:2.2}
\ee
The first Noether theorem implies
\be
u_\al u^\al=k_1,~~~                  
u_\al \fr{Dr^\al}{d\ta}=k_2.
\label{eq:2.3}
\ee
Varying the action
\be
2\de W=\int^{\ta_1}_{\ta_0}\left[(\De u_\al)\fr{Dr^\al}{d\ta}
      +u_\al\De\left(\fr{Dr^\al}{d\ta}\right)\right]d\ta,
\label{eq:2.4}
\ee
then using
\be
\De u^\al=\fr{D}{d\ta}\de x^\al,~~~
\De \fr{Dr^\al}{d\ta}=\fr{D\De}{d\ta}r^\al
      +R^\al_{~\la\mu\nu}r^\la\de x^\mu u^\nu,
\label{eq:2.5}
\ee                      
gives
\ber
2\de W&=&\int^{\ta_1}_{\ta_0}d\ta
         \left[\fr{D\de x_\al}{d\ta}\fr{Dr^\al}{d\ta}
               +u_\al\fr{D}{d\ta}\De r^\al
               +u_\al R^\al_{~\la\mu\nu}r^\la\de x^\mu u^\nu\right]\nonumber\\
&=&\int^{\ta_1}_{\ta_0}d\ta\left(\left[\fr{D^2r_\al}{d\ta^2}  
       +u_\ga R^\ga_{~\la\mu\nu}r^\la u^\nu\right]\de x^\al  
     +\fr{Du_\al}{d\ta}\De r^\al\right)\nonumber\\ 
&+&\left(u_\al\De r^\al
     +\fr{Dr^\al}{d\ta}\de x_\al\right)|^{\ta_1}_{\ta_0}.
\label{eq:2.6}
\ear
Thus 
\be
\fr{2\de W}{\De r^\al}=\fr{Du_\al}{d\ta},
\label{eq:2.7}
\ee
and
\be
\fr{2\de W}{\de x_\al}=\fr{D^2 r^\al}{d\ta^2}
                        -R^\al_{~\bt\ga\de}u^\bt R^\ga u^\de.
\label{eq:2.8}
\ee
Taking these variations to vanish gives the geodesic equation and the geodesic
deviation equation respectively.   The conjugate Lagrangian momenta are
\be
p_\al=\fr{\p L}{\p r^\al}=u_\al,~~~     
\pi_\al=\fr{\p L}{\p u^\al}=\fr{D r_\al}{d\ta}.
\label{eq:2.9}
\ee
Using these the first integrals derived from the first Noether theorem can 
be rewritten as
\be
\si_1\equiv p_\al p^\al-m^2=0,~~~
\si_2\equiv p_\al \pi^\al-l^2=0,
\label{eq:2.10}
\ee
where $k_1=m^2$ and $k_2=l^2$.   These expressions for the momenta and first 
integrals suggest the extended phase space action given in the next section.
\section{The Phase Space Action.}
\label{sec:3}
The extended phase space action which generalizes the action of 
Pavsic (1987) \cite{bi:pavsic} for geodesics is
\be
I=\int^{\ta_1}_{\ta_0}d\ta\left[p_\al\fr{Dr^\al}{d\ta}
       +u_\al\pi^\al+\la_1(p_\al p^\al-m^2)+\la_2(p_\al\pi^\al-l^2)\right].
\label{eq:3.1}
\ee
Varying with respect to $\la_1, \la_2, \pi^\al,  p^\al$ is straightforward
\ber
\fr{\de I}{\de \la_1}&=&p_\al p^\al-m^2,~~~    
\fr{\de I}{\de \la_2}=p_\al\pi^\al-l^2,\nonumber\\
\fr{\de I}{\de \pi^\al}&=&u_\al+\la_2p_\al,~~~
\fr{\de I}{\de p_\al}=\fr{Dr^\al}{d\ta}+2\la_1 p^\al+\la_2\pi^\al.
\label{eq:3.2}
\ear
Varying with respect to $r^\al$ and $x^\al$
\be
\de I_{(rem.)}=\int^{\ta_1}_{\ta_0}d\ta\left[p_\al\De\fr{Dr^\al}{d\ta}
                                        +\pi^\al\De u^\al\right],
\label{eq:3.3}
\ee
again using the expressions \ref{eq:2.5} gives
\ber
\de I_{(rem.)}&=&\int^{\ta_1}_{\ta_)}d\ta\left[p_\al\fr{D\De r^\al}{d\ta}
                 +p_\ga R^\ga_{\la\al\nu}r^\la\de x^\al u^\nu
                 +\pi_\al\fr{D}{d\ta}\de x^\al\right]\nonumber\\
              &=&\int^{\ta_1}_{\ta_0}d\ta\left(\left[\fr{D\pi_\al}{d_ta}
                 +p_\ga R^\ga_{~\la\al\nu}r^\la u^\nu\right]\de x^al
                   +\fr{Dp_\al}{d\ta}\De r^\al\right)\nonumber\\
     && \quad +\left(p_\al\De r^\al+\pi_\al\de x^\al\right)|^{\ta_1}_{\ta_0}.
\label{eq:3.4}
\ear
Thus the remaining variations give
\be
\fr{\de I}{\de r^\al}=\fr{D p_\al}{d\ta},~~~      
\fr{\de I}{\de x^\al}=\fr{d\pi_\al}{d\ta}-R_{\al\bt\ga\de}u^\bt r^\ga p^\de.
\label{eq:3.5}
\ee
Setting these variations \ref{eq:3.2} and \ref{eq:3.5} equal to zero gives 
six equations,  the first four can be used to give expressions for
$\la_1, \la_2, \pi_\al,  p_\al$ thus:
\ber
\la_1&=&\pm\fr{\sqrt{u_\ga u^\ga}}{2m}\left(\fr{u_\ga\dot{r}^\ga}{u_\ga u^\ga}
                                        -\fr{l^2}{m^2}\right),~~~~~
\la_2=\pm\fr{\sqrt{u_\ga u^\ga}}{m},\nonumber\\
\pi^\al&=&\mp\fr{m}{\sqrt{u_\ga u^\ga}}\left(h^\al_{~\bt}\dot{r}^\bt
                                        +\fr{l^2}{m^2}u^\al\right),~~~
p^\al=\mp\fr{m u^\al}{\sqrt{u_\ga u^\ga}},
\label{eq:3.6}
\ear
where $h_{\al\bt}$ is the projection tensor
\be
h_{\al\bt}=g_{\al\bt}-\fr{u_\al u_\bt}{u_\ga^\ga},
\label{eq:3.7}
\ee
and the $\pm$ arises from taking the square root when solving for $\la_2$,  
using the $\de \pi_\al$ and $\de\la_1$ equations,  the lower sign should be 
used if the momentum and velocity are to be co-directional.   
These four equations \ref{eq:3.6} can be substituted into 
the two remaining equations \ref{eq:3.5} to give the general 
non-normalized geodesic and geodesic deviation equations in the form
of Bazanski (1977) \cite{bi:baz77} eq.1.6 and 2.4
\ber
&&\fr{D}{d\ta}\fr{mu_\al}{\sqrt{u_gau^\ga}}=0,\nonumber\\  
&&\fr{D}{d\ta}\left(\fr{h_{\al\bt}}{\sqrt{u_\ga u^\ga}}\fr{Dr^\bt}{d\ta}\right)
   -\fr{1}{\sqrt{u_gau^\ga}}R_{\al\bt\ga\de}u^\bt r^\ga u^\de=0,
\label{eq:3.8}
\ear
The Poisson brackets for variables $A$ and $B$ at equal $\ta$ are defined by
\be
\vb+{+A,B\vb+}+=\fr{\De A}{\p r_\ga}\fr{\De B}{\p p^\ga}
               -\fr{\De A}{\p p_\ga}\fr{\De B}{\p r^\ga}   
               +\fr{\De A}{\p x_\ga}\fr{\De B}{\p \pi^\ga}
               -\fr{\De A}{\p \pi_\ga}\fr{\De B}{\p x^\ga},
\label{eq:3.9}
\ee
where the $\na/\p x^\ga$ signifies that covariant derivatives must be taken.
The total Hamiltonian is
\be
H=\mp\fr{\sqrt{u_\ga u^\ga}}{2m}\left[\left(\fr{u_\al \dot{r}^\al}{u_\ga u^\ga}
  -\fr{l^2}{m^2}\right)(p_\bt p^\bt-m^2)+2(p_\al\pi^\al-l^2)\right],
\label{eq:3.10}
\ee
Taking the two $\la$ variations of the extended phase space action \ref{eq:3.1}
to vanish implies that the two terms in the total Hamilton vanish separately.
The Hamiltonian equations of motion are
\ber             
\dot{x}_\al&=&\vb+{+x_\al,H\vb+}+=\fr{\De H}{\p \pi^\al},~~~
\dot{p}_\al=\vb+{+p_\al,H\vb+}+=-\fr{\De H}{\p r^\al},\nonumber\\
\dot{r}_\al&=&\vb+{+r_\al,H\vb+}+=\fr{\De H}{\p p^\al},~~~
\dot{\pi}_\al=\vb+{+\pi_\al,H\vb+}+=-\fr{\De H}{\p x^\al},
\label{eq:3.11}
\ear
where the dot as in $\dot{p}_\al$,  again signifies that covariant derivatives
are taken,  for example $\dot{\pi}_\al=D\pi_al/d\ta$.   
The last equation is proved here,  the others are similar and simpler.
$\de/\p x^\al$ acting on $u^\al$ and $p^\al$ vanishes,
thus the equation reduces to 
$\dot{\pi}_\al=\mp u_\ga u^\ga m ^{-1}\De \pi_\al/\p x^\bt p^\bt$,
using the expression for $p_\al$ and that 
$u_\bt\pi^\al_{~;\bt}=D\pi_\al/d\ta$ gives    
$\dot{\pi}_\al=D\pi_\al/d\ta$.   Conjugate $(p,x)$ and $(\pi,r)$ 
in the Poisson bracket the Hamiltonian equations of motion are not recovered,
as can be seen immediately from the equation for $\dot{x}^\al$.
\section{Quantization.}
\label{sec:4}
The gauges could be fixed by introducing
\be
\si_3=x^t-\ta=0,~~~
\si_4=p_\al r^\al=0,
\label{eq:4.1}
\ee
and then calculating $C_{\al\bt}=\vb+{+\si_al,\si_\bt\vb+}+$  
in order to produce Dirac brackets,  but here just Poisson brackets are used.
The coordinates and momenta in phase space obey
\be
\vb+{+p_\al,r^\bt\vb+}+=\vb+{+\pi_\al,x^\bt\vb+}+=-\de_\al^{~\bt}.
\label{eq:4.2}
\ee
To quantize the system the Poisson brackets are replaced by Heisenberg 
brackets,  applied to \ref{eq:4.2} this implies the operator substitutions
\be
p_\al\rightarrow-\fr{i\hb\na}{\p r^\al},~~~
\pi_\al\rightarrow-\fr{i\hb\na}{\p x^\al}.
\label{eq:4.3}
\ee
Assuming that both terms in the total Hamiltonian vanish separately these 
operator substitutions give
\be
\left(\fr{\na}{\p r^\al}\fr{\na}{\p r_\al}+\fr{m^2}{\hb^2}\right)\ps=0,~~~
\left(\fr{\na}{\p r^\al}\fr{\na}{\p x_al}+\fr{l^2}{\hb^2}\right)\ps=0,
\label{eq:4.4}
\ee
where the wave function $\ps$ is dependent on both $x^\al$ and $r^\al$.   
The operator substitutions are not applied to equation \ref{eq:3.5},  
for example in the quantization of just a free particle 
the geodesic equation can be written as
$u^\bt p_{\al;\bt}=0$ or $p^\bt p_{\al;\bt}=0$,   
and the operator substitutions would give
$u^\bt\ps_{\al\bt}=0$ or $\ps_\bt R^\bt_{~\al}+m^2\ps_\al=0$,   
the first of these gives an unusual restriction on the wave function,  
the flat space limit shows that the second of these is incorrect;   
because the operator substitutions are not applied to \ref{eq:3.5} 
the Riemann tensor does not explicitly occur in the differential 
equation for the wave function \ref{eq:4.4}.   Taking
\be
\ps=A~exp(\fr{i}{\hb}S),
\label{eq:4.5}
\ee
and multiplying by $\hb^2/\ps$,  \ref{eq:4.4} becomes
\ber
i\hb S_{r_\al}^{~~~r_\al}-S_{r_\al}S^{r_\al}+m^2=0,\nonumber\\
i\hb S_{r_\al}^{~~~x_\al}-S_{r_\al}S^{x_\al}+l^2=0,
\label{eq:4.6}
\ear
noting from ref{eq:3.4} that the principle Hamiltonian $S$ obeys 
$p_\al=\na S/\p r^\al$ and 
$\pi_\al=\na S/\p x^\al$ shows that $\si_1$ and $\si_2$    
in equation \ref{eq:2.10} are recovered in the limit $\hb\rightarrow 0$.   
Defining
\be
S\equiv r^\ga U_\ga+V,
\label{eq:4.7}
\ee
where $U$ and $V$ are functions of only $x^\al$,  shows that \ref{eq:4.6} 
can be written in the form
\be
u_\al u^\al=m^2,~~~
-i\hb U_\al^{~\al}+U_\al V^\al=l^2.
\label{eq:4.8}
\ee
No general separation of the wave function $\ps$ into $x^\al$ and $r^\al$
dependent parts is known;   however a particular case is obtained by defining
\be
\ps=A~exp(\fr{r^\bt r^\ga\ph_{\bt\ga}}{\ph}),
\label{eq:4.9}
\ee
where $\ph$ is a function of $x^\al$ only,  
then using the quantum analog of $p_\al r^\al=0$ given by 
\be
r^\al\ph_\al=a,
\label{eq:4.10}
\ee
where $a$ is a constant,  and $l=0$,  the first equation in \ref{eq:4.4} 
reduces to the Klein-Gordon equation
\be
\ph^\al_{~\al}+\fr{m^2}{\hb^2}\ph=0,
\label{eq:4.11}
\ee
and the second vanishes identically.

\end{document}